\documentstyle[12pt]{article}
\def\doublespace{\def\baselinestretch{1.3}\Large\normalsize}
\title{\bf A Supersymmetric Treatment of a Particle Subjected to a Ring-shaped
Potential}
\author{ GARDO G. BLADO\\
         \it Division of Science and Mathematics
         University of Minnesota, Morris\\
            \it Morris, MN, U.S.A. 56267-2128}
\date{ }
\begin{document}
\doublespace
\maketitle
\begin{abstract}
The ring-shaped Hartmann potential
$
V = \eta \sigma^{2} \epsilon_{0} \left( \frac{2 a_{0}}{r} - \frac{\eta
a_{0}^{2}}{r^{2} sin^{2} \theta}    \right)
$
was introduced in quantum chemistry to describe ring-shaped molecules like
benzene. In this article, fundamental concepts
of supersymmetric quantum mechanics (SUSYQM) are discussed.
 The energy eigenvalues and
(radial) eigenfunctions  of the Hartmann potential are subsequently rederived
using the
techniques of SUSYQM.
\vspace{2.5in}
\begin{tabbing}
           \> ring-shaped potential, superpotential
\end{tabbing}
\end{abstract}
\clearpage

\section{Introduction}
\label{intro}
\indent

In 1972, an exactly solvable ring-shaped potential was introduced by H.
Hartmann \cite{hart}. The Hartmann potential is given by
the following expression,
\begin{equation}
V = \eta \sigma^{2} \epsilon_{0} \left( \frac{2 a_{0}}{r} - \frac{\eta
a_{0}^{2}}{r^{2} sin^{2} \theta}    \right)
\label{1susyhh1}
\end{equation}
where
\begin{equation}
 a_{0} = \frac{\hbar^{2}}{\mu e^{2}}\;\;\;\;\;\;\;\; \rm and \;\;\;\;\;\;\;\;
\mit \epsilon_{0} = -\frac{1}{2} \frac{\mu e^{4}}{\hbar^{2}}
\end{equation}
$\mu$ is the particle mass, $\eta$ and $\sigma$ are positive real parameters
which range from about 1 to 10 in theoretical chemistry applications
\cite{schuch} and $r$, $\theta$ are in spherical coordinates. Following the
exact solution of the Schr\"{o}dinger equation given by Hartmann
\cite{hart}, alternative methods of solutions in spherical coordinates
\cite{sok} and squared parabolic coordinates \cite{filho,chet,gerry,kibler,neg}
had been given. In this article, the author presents another alternative method
of solution in spherical coordinates using supersymmetry (SUSY).

The concept of supersymmetry (SUSY) has been used in particle physics in the
past two decades \cite{nil,haber}. It was discovered in 1971
by Gel'fand and Likhtman \cite{gelfand}. Simply put, supersymmetry is a
symmetry which relates fermionic and bosonic degrees
of freedom. At present, particle physicists believe that it is an essential
ingredient in unifying the four fundamental forces
in nature namely the electromagnetic, weak, strong and gravitational
interactions.

Supersymmetric theories of the four fundamental interactions entail the
presence of SUSY partners which have the same mass as their
corresponding ordinarily observed particles.\footnote{For example, the SUSY
partner of the electron is called a selectron while that of
a photon is a photino.}
Unfortunately, these have not
been observed in nature. To make sense out of this experimental fact, theorists
believe that SUSY must be ``broken" at ordinary
energies. The search for a mechanism to break SUSY led Witten \cite{witten} in
1981 to study SUSY breaking in the simplest
case of SUSY quantum mechanics. In fact, studies in SUSYQM during its early
years, were confined solely for understanding SUSY
breaking.

However, it was eventually discovered that SUSYQM can have interesting
applications besides its use in the study of SUSY
breaking. At present, it has found its way in many areas of physics including
atomic physics, statistical physics, nuclear
physics. etc. \cite{cooper}. Through the present article, the author hopes to
contribute to the utilization of SUSYQM in
theoretical chemistry.

One simple use of SUSYQM is in obtaining the eigenvalues and eigenfunctions of
the Schr\"{o}dinger equation \cite{schwabl,genden}.
 A two-dimensional supersymmetric solution of the Hartmann potential in
``square" parabolic coordinates \cite{filho}
had been proposed. In the present article, the author formulates a
one-dimensional SUSY solution similar to that of the hydrogen atom
\cite{schwabl}. This avoids the complications of a two-dimensional SUSYQM
formulation. Using SUSY, it is shown that the eigenvalues
and (radial) eigenfunctions  can be obtained. The techniques employed here are
largely due to
Schwabl \cite{schwabl}.

In section \ref{susyqm}, a pedagogical introduction to SUSYQM is developed.
Only the concepts and equations which are essential
for the present paper are presented.

Section \ref{susyhh} gives a detailed account of how SUSYQM can be used to
obtain the eigenvalues and (radial) eigenfunctions of the
Hartmann potential. The discussion is heavily supplemented by energy level
diagrams to facilitate the understanding of the hierarchy of
hamiltonians and how the SUSY operators $A^{+}_{L}$ are used to obtain the
eigenfunctions.

Some conclusions are given in section \ref{concl}.
\clearpage

\section{Supersymmetric Quantum Mechanics}
\label{susyqm}
\indent

As mentioned in the Introduction, SUSY was first applied to particle physics,
whose language is quantum field theory. In quantum field theory,
a particle is represented by a component field $\varphi_{i}$ and its dynamics
is described by a lagrangian density
$\cal L \mit (\varphi_{i}, \partial_{\mu} \varphi_{i})$ where $\left[
\partial_{\mu} \equiv \left( \frac{1}{c} \frac{\partial}{\partial t},
\vec{\nabla}\right) \right]$. The word ``supersymmetry" was originally used to
describe the symmetry which transforms a field $\varphi$ to another
field $\psi$ whose intrinsic spin differs from $\varphi$ by $\frac{1}{2}
\hbar$. In SUSYQM, which will be described here, we will use the term
``supersymmetry" in a more general sense. It will be used to denote systems
which can be described by the SUSY algebra in supersymmetric field theory.

SUSYQM \cite{witten,suku} is characterized by the existence of the charge
operators $Q_{i}$, where $i = 1, 2, \ldots, N$ such that they obey the SUSY
algebra
(denoted by sqm($N$)),
\begin{equation}
\left\{  Q_{i}, Q_{j} \right\} = \delta_{ij} H_{ss}\;\;\;\;\;\;\;\; \left[
Q_{i}, H_{ss} \right] = 0
\label{1susyqm1}
\end{equation}
where $H_{ss}$ is the supersymmetric Hamiltonian, \{ \} and [ ] are
anticommutator and commutator respectively. We consider only sqm(2)
with charge operators $Q_{1}$ and $Q_{2}$ and construct the linear combinations
\begin{equation}
Q = \frac{1}{\sqrt{2}} \left( Q_{1} + i Q_{2}  \right) \;\;\;\;\;\;\;\; \rm and
\;\;\;\;\;\;\;\; \mit Q^{\dagger} =
\frac{1}{\sqrt{2}} \left( Q_{1} - i Q_{2}  \right)\;.
\label{2susyqm1}
\end{equation}
{}From equations \ref{1susyqm1}  and \ref{2susyqm1}, The SUSY algebra is then
\begin{equation}
\left\{  Q, Q^{\dagger} \right\} = H_{ss}, \;\;\;\;\;\; Q^{2} = 0, \;\;\;\;\;\;
\left( Q^{\dagger}  \right)^{2} = 0
\label{3susyqm1}
\end{equation}
with
\begin{equation}
\left[  Q, H_{ss} \right] = 0 \;\;\;\;\;\;\;\; \rm and \;\;\;\;\;\;\;\; \mit
\left[  Q^{\dagger}, H_{ss} \right] = \rm 0.
\end{equation}

The above SUSY algebra can be realized by letting
\begin{equation}
Q = \left[ \begin{array}{cc}
0 &  0    \\
 A^{-} & 0
\end{array} \right]
\;\;;\;\;
Q^{\dagger} = \left[ \begin{array}{cc}
0 & A^{+}     \\
0 & 0
\end{array} \right]
\label{2susyqm2}
\end{equation}
where
\begin{equation}
\left(  A^{-} \right)^{\dagger} = A^{+}.
\end{equation}
{}From equations \ref{3susyqm1} and \ref{2susyqm2}, the supersymmetric
hamiltonian is
\begin{equation}
H_{ss} = \left[ \begin{array}{cc}
H_{1} &  0    \\
 0 & H_{2}
\end{array} \right]
\label{3susyqm2}
\end{equation}
where
\begin{equation}
H_{1} = A^{+} A^{-}\;\;\;\;\;\;\;\; \rm and \;\;\;\;\;\;\;\; \mit H_{2} = A^{-}
A^{+}.
\label{4susyqm2}
\end{equation}
The hamiltonians $H_{1}$ and $H_{2}$ are said to be ``supersymmetric" partners
of each other.

The hamiltonian of the Schr\"{o}dinger equation can always be factorized in the
form of equation \ref{4susyqm2}. Consider the hamiltonian
\cite{cooper}
\begin{equation}
H_{1} = -\frac{1}{2} \frac{d^{2}}{dx^{2}} + V_{1}(x)
\label{5susyqm2}
\end{equation}
such that
\begin{equation}
H_{1} \psi^{n}_{(1)} = \left[ -\frac{1}{2} \frac{d^{2}}{dx^{2}} + V_{1}(x)
\right] \psi^{n}_{(1)} = E^{n}_{(1)} \psi^{n}_{(1)}
\label{5psusyqm2}
\end{equation}
where $V_{1}(x)$ is chosen such that the ground state $\psi^{0}_{(1)}$ has
energy equal to zero. The hamiltonian in equation
\ref{5susyqm2} can be put in the form of equation \ref{4susyqm2} by letting
\begin{equation}
A^{-}_{1} = \frac{1}{\sqrt{2}} \left( \frac{d}{dx} + W_{1}
\right)\;\;\;\;\;\;\;\; \rm and \;\;\;\;\;\;\;\; \mit
A^{+}_{\rm 1} = \frac{\rm 1}{\sqrt{\rm 2}} \left( -\frac{d}{dx} + W_{\rm 1}
\right)
\label{6susyqm2}
\end{equation}
provided that the ``superpotential'' $W_{1}$ satisfies the Ricatti equation
\begin{equation}
V_{1}(x) = \frac{1}{2} \left[  W_{1}^{2} - \frac{dW_{1}}{dx}\right]\;.
\label{1susyqm3}
\end{equation}
As long as equation \ref{1susyqm3} has a solution $W_{1}$, the one-dimensional
Schr\"{o}dinger equation can be made supersymmetric
by the construction given in equations \ref{6susyqm2}, \ref{4susyqm2} and
\ref{3susyqm2}. The challenge then in using SUSYQM techniques
is not in the mechanics of the construction of just any SUSY hamiltonian, but
in finding a suitable superpotential (or $V_{1}(x)$) to
construct a SUSY hamiltonian which will be relevant to the problem at hand. It
is a common practice to choose or pose as an ansatz the
$W_{1}$ to solve a physical problem \cite{schwabl,rico}.

The SUSY partner of $H_{1}$, namely $H_{2}$ is then given by
\begin{equation}
H_{2} = -\frac{1}{2} \frac{d^{2}}{dx^{2}} + V_{2}(x) = A^{-}_{1} A^{+}_{1}
\label{2susyqm3}
\end{equation}
where
\begin{equation}
V_{2}(x) = \frac{1}{2} \left[  W_{1}^{2} + \frac{dW_{1}}{dx}\right]\;.
\end{equation}

Note that $H_{2}$ is altogether a new hamiltonian. An astute reader will
immediately realize that one can repeat the procedure in constructing $H_{2}$
from $H_{1}$ to construct an $H_{3}$ from $H_{2}$ such that
\begin{equation}
H_{2} = A^{+}_{2} A^{-}_{2}
\end{equation}
with
\begin{equation}
A^{-}_{2} = \frac{1}{\sqrt{2}} \left( \frac{d}{dx} + W_{2}
\right)\;\;\;\;\;\;\;\; \rm and \;\;\;\;\;\;\;\; \mit
A^{+}_{\rm 2} = \frac{\rm 1}{\sqrt{\rm 2}} \left( -\frac{d}{dx} + W_{\rm 2}
\right)
\end{equation}
and with a new Ricatti equation
\begin{equation}
V_{2}(x) = \frac{1}{2} \left[  W_{2}^{2} - \frac{dW_{2}}{dx}\right].
\label{1susyqm4}
\end{equation}
$W_{2}$ in equation \ref{1susyqm4} is then solved to construct $A^{\pm}_{2}$.
$H_{3}$ can then be constructed as
\begin{equation}
H_{3} = -\frac{1}{2} \frac{d^{2}}{dx^{2}} + V_{3}(x) = A^{-}_{2} A^{+}_{2}
\end{equation}
where
\begin{equation}
V_{3}(x) = \frac{1}{2} \left[  W_{2}^{2} + \frac{dW_{2}}{dx}\right]\;.
\end{equation}
We can evidently construct a ``hierarchy" of SUSY-partner hamiltonians $H_{1}$,
$H_{2}$, $H_{3}$, \ldots , $H_{n}$ starting from $H_{1}$.

Let us go back to the first two hamiltonians we started with namely $H_{1}$ and
$H_{2}$. Since $V_{1}(x)$ in equation \ref{5susyqm2} is chosen
such that its ground state wave function $\psi^{0}_{(1)}$ has eigenvalue equal
to zero, equation \ref{4susyqm2} gives
\begin{equation}
H_{1} \psi^{0}_{(1)} = 0 \;\;\; \Longrightarrow \;\;\; A^{+}_{1} A^{-}_{1}
\psi^{0}_{(1)} = 0.
\label{4susyqm4}
\end{equation}
Equation \ref{4susyqm4} implies
\begin{equation}
A^{-}_{1} \psi^{0}_{(1)} = 0.
\label{5susyqm4}
\end{equation}
With equation \ref{6susyqm2} and knowing $W_{1}$, equation \ref{5susyqm4}
allows one to calculate the ground state of $H_{1}$ by solving the
resulting first order differential equation.

Consider any eigenstate of $H_{1}$, $\psi^{n}_{(1)}$ with energy $E^{n}_{(1)}$.
We have
\begin{equation}
H_{1} \psi^{n}_{(1)} = E^{n}_{(1)}\psi^{n}_{(1)}
\end{equation}
or from equation \ref{4susyqm2}
\begin{equation}
A^{+}_{1} A^{-}_{1} \psi^{n}_{(1)} = E^{n}_{(1)} \psi^{n}_{(1)}.
\label{1susyqm5}
\end{equation}
Applying $A^{-}_{1}$ to equation \ref{1susyqm5},
$A^{-}_{1} A^{+}_{1} \left( A^{-}_{1} \psi^{n}_{(1)}  \right) = E^{n}_{(1)}
A^{-}_{1} \psi^{n}_{(1)}$
or with equation \ref{2susyqm3}
\begin{equation}
H_{2} \left( A^{-}_{1} \psi^{n}_{(1)}  \right) = E^{n}_{(1)} \left( A^{-}_{1}
\psi^{n}_{(1)} \right).
\label{2susyqm5}
\end{equation}

Conversely, consider an eigenfunction $\psi^{n}_{(2)}$ of $H_{2}$ with
eigenvalue $E^{n}_{(2)}$. With equation \ref{2susyqm3}, we get
$H_{2} \psi^{n}_{(2)} = A^{-}_{1} A^{+}_{1} \psi^{n}_{(2)} = E^{n}_{(2)}
\psi^{n}_{(2)}$. Multiplying by $A^{+}_{1}$, we have,
$A^{+}_{1} A^{-}_{1} \left( A^{+}_{1} \psi^{n}_{(2)}  \right) = E^{n}_{(2)}
A^{+}_{1} \psi^{n}_{(2)}$. With equation \ref{4susyqm2}
\begin{equation}
H_{1} \left( A^{+}_{1} \psi^{n}_{(2)}  \right) = E^{n}_{(2)} \left( A^{+}_{1}
\psi^{n}_{(2)} \right).
\label{3susyqm5}
\end{equation}

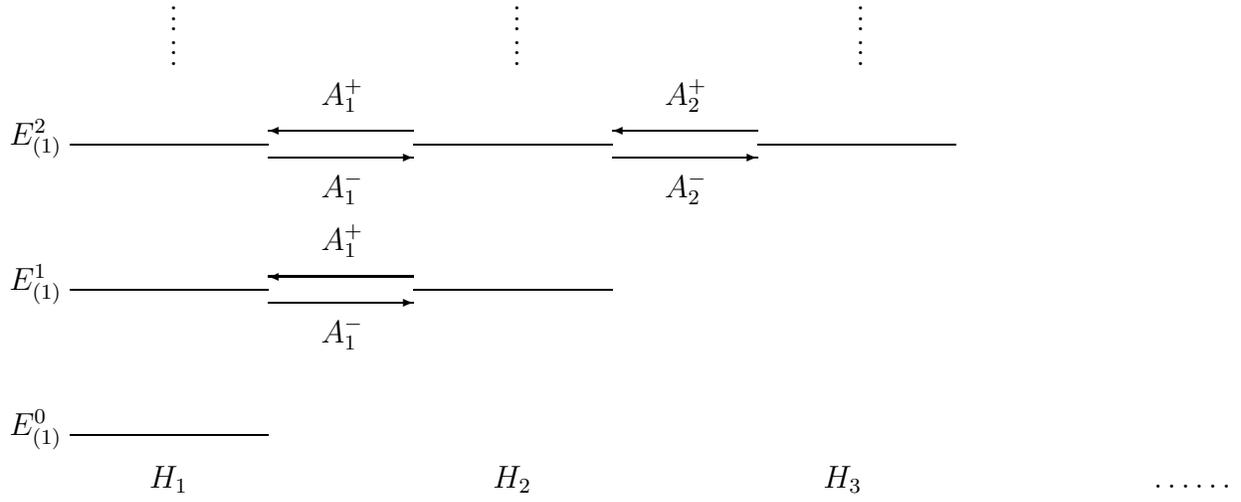
\begin{figure}[htbp]
\begin{center}
\begin{picture}(400,200)
                            \put(37.5,173){\vdots}
        \put(167.5,173){\vdots}
\put(297.5,173){\vdots}
                            \put(37.5,160){\vdots}
        \put(167.5,160){\vdots}
\put(297.5,160){\vdots}

\put(95,145){$A^{+}_{1}$}
\put(225,145){$A^{+}_{2}$}

\put(130,135){\vector(-1,0){55}}
\put(260,135){\vector(-1,0){55}}
\put(-23,130){$E^{2}_{(1)}$} \put(0,130){\line(75,0){75}}
         \put(130,130){\line(75,0){75}}
\put(260,130){\line(75,0){75}}

\put(75,125){\vector(1,0){55}}
\put(205,125){\vector(1,0){55}}

\put(95,110){$A^{-}_{1}$}
\put(225,110){$A^{-}_{2}$}

\put(95,90){$A^{+}_{1}$}

\put(130,80){\vector(-1,0){55}}
\put(-23,75){$E^{1}_{(1)}$} \put(0,75){\line(75,0){75}}
        \put (130,75){\line(75,0){75}}

\put(75,70){\vector(1,0){55}}

\put(95,55){$A^{-}_{1}$}
\put(-23,20){$E^{0}_{(1)}$} \put(0,20){\line(75,0){75}}
                            \put(30,0){$H_{1}$}
        \put(160,0){$H_{2}$}
\put(285,0){$H_{3}$}           \put(410,0){\ldots \ldots}
\end{picture}
\end{center}
\caption{The hierarchy of hamiltonians and the action of the operators
$A^{\pm}_{n}$ on the degenerate eigenstates}
\label{fig1susyqm5}
\end{figure}

Equations \ref{2susyqm5} and \ref{3susyqm5} imply that the hamiltonians $H_{1}$
and $H_{2}$ have identical eigenvalues except for the ground state
$\psi^{0}_{(1)}$ of $H_{1}$ (since $A^{-}_{1} \psi^{0}_{(1)}=0$ in equation
\ref{2susyqm5} and this is unnormalizable). In addition, we  can see that
if you know an eigenfunction of $H_{1}$, i.e. $\psi^{n}_{(1)}$, then an
eigenfunction $A^{-}_{1} \psi^{n}_{(1)}$ of $H_{2}$ can be formed. Similarly,
an eigenfunction $A^{+}_{1} \psi^{n}_{(2)}$ of $H_{1}$ can be formed given an
eigenfunction $\psi^{n}_{(2)}$ of $H_{2}$. The preceding analysis can
then be extended to the hierarchy of hamiltonians discussed earlier. These
observations are illustrated in figure \ref{fig1susyqm5}
\cite{schwabl,suku}.

Herein lies a very important consequence of SUSYQM. The energy eigenfunctions
of the hierarchy of hamiltonians are related by the
SUSY operators $A^{\pm}$. If one knows the eigenvalues and eigenfunctions of a
particular $H_{n}$, then  one can get the eigenvalues and eigenfunctions of its
SUSY
partner.

Note that what we have discussed is SUSYQM in one dimension.There had been work
done in doing SUSYQM in two or more dimensions \cite{filho,andria}. To be able
to apply
one dimensional SUSYQM to the Hartmann potential, we will do a separation of
variables of the resulting Schr\"{o}dinger equation. To be able to form a
hamiltonian
from
 a separated one dimensional
differential equation that can be an element of an $H_{ss}$, this one
dimensional differential equation
 must be of the form of equation \ref{5psusyqm2} (no first derivative term) and
must yield an infinite tower of
states as for $H_{n}$ of figure \ref{fig1susyqm5} \cite{bluhm}. As we will see,
upon separation of variables for the Hartmann potential, only the radial
equation yields
an interesting SUSY.
\clearpage

\section{Supersymmetry in the Hartmann Hamiltonian}
\label{susyhh}
\indent

  With the SUSYQM concepts introduced in section \ref{susyqm}, we are now ready
to
obtain the eigenvalues and radial eigenfunctions of the Hartmann potential
in spherical coordinates.

The Schr\"{o}dinger equation in spherical coordinates for a particle of mass
$\mu$ subjected to the Hartmann potential in equation \ref{1susyhh1}
is given by
\begin{equation}
-\frac{\hbar^{2}}{2 \mu} \nabla^{2} \psi + \left[ \frac{2 \eta \sigma^{2}
\epsilon_{0} a_{0}}{r}  -
\frac{\eta^{2} \sigma^{2} a_{0}^{2} \epsilon_{0}}{r^{2} sin^{2} \theta}
\right] \psi = E \psi\;.
\label{3susyhh1}
\end{equation}
Assuming a solution
\begin{equation}
\psi = R(r) \Theta(\theta) \Phi(\phi)
\end{equation}
equation \ref{3susyhh1} can be separated into three differential equations
\cite{hart}
\begin{equation}
\frac{1}{\Phi} \frac{d^{2} \Phi}{d \phi^{2}} = - m^{2}
\label{1susyhh2}
\end{equation}
\begin{equation}
\frac{1}{sin \theta} \frac{d}{d \theta} \left( sin \theta \frac{d \Theta}{d
\theta} \right) -
\left( \frac{M^2}{sin^2 \theta} - L(L+1) \right) \Theta = 0
\label{2susyhh2}
\end{equation}
\begin{equation}
\frac{1}{r^2} \frac{d}{dr} \left( r^2 \frac{dR}{dr} \right) - L(L+1)
\frac{R}{r^2} +
\frac{8 \pi^2 \mu}{h^2} \left( E + \frac{\eta \sigma^2 e^2}{r} \right) R = 0
\label{3susyhh2}
\end{equation}
where
\begin{equation}
M^2 = m^2 + \eta^2 \sigma^2 \;.
\end{equation}

Looking at equations \ref{1susyhh2} to \ref{3susyhh2}, we realize that these
closely resemble the separated equations of the hydrogen atom \cite{grif}.
As shown by reference \cite{bluhm}, the only interesting separable SUSY in the
hydrogen atom in spherical coordinates results from the radial equation.
Their argument is as follows. Looking at equation \ref{1susyhh2}, and comparing
it with equation \ref{5psusyqm2}, we see that $V_{1} = 0$. Equation
\ref{2susyhh2} on the other hand can be cast to a form similar to equation
\ref{5psusyqm2} by multiplying it by a modulation factor
$\frac{f(cos \theta)}{\left[ 1 - cos^2 \theta  \right]^{1/2}}$. The eigenvalue
of $f(cos \theta)$  is zero and no infinite tower of states can be generated.
Hence,
the $\Phi$ and $\Theta$ solutions cannot be given by SUSYQM. They can be solved
by conventional means \cite{hart}
\begin{equation}
\Phi(\phi) = \frac{1}{\sqrt{2 \pi}} e^{i m \phi}, \;\;\;\;\; m = 0, \;\pm 1,\;
\pm 2, \ldots
\end{equation}
\begin{equation}
\Theta (\theta) = \cal P ^{\left| \mit M \right|}_{\mit L} (\mit cos \theta),
\;\;\;\;\; \mit L = \nu^{\prime} + \left| \mit M \right| ,
\;\;\;\;\; \nu^{\prime} = \rm 0,\;1,\;2, \ldots
\label{6susyhh2}
\end{equation}
where $\cal P ^{\left| \mit M \right|}_{\mit L} (\mit cos \theta)$ reduces to
the associated Legendre polynomials  when $\eta \sigma \longrightarrow 0$.

The radial equation \ref{3susyhh2} can be cast into a form similar to
\ref{5psusyqm2} by letting
\begin{equation}
R = \frac{u}{r}\;.
\label{1susyhh3}
\end{equation}
Substituting equation \ref{1susyhh3} into equation \ref{3susyhh2}, yields
\begin{equation}
H_{L} u = \left[ -\frac{1}{2} \frac{d^2}{dr^2} +  \frac{L(L+1)}{2 r^2} -
\frac{\gamma}{r} \right] u = \frac{\mu E}{\hbar^2} u
\label{2susyhh3}
\end{equation}
with
\begin{equation}
\gamma \equiv \frac{\mu \eta \sigma^2 e^2}{\hbar^2}.
\end{equation}

Equation \ref{2susyhh3} is similar to that of the hydrogen atom's radial
equation. We thus claim that we can obtain the eigenvalues and radial
eigenfunctions by looking at the hamiltonian \cite{schwabl}
\begin{equation}
\cal H_{\mit L} = \mit -\frac{\rm 1}{\rm 2} \frac{d^2}{dr^2} + \frac{L(L+\rm
1)}{\rm 2 \mit r^{2}} - \frac{\gamma}{r} + \frac{\rm 1}{\rm 2} \left(
\frac{\gamma}{L+\rm 1}  \right)^2
\label{4susyhh3}
\end{equation}
which yields a Ricatti equation (from equations \ref{4susyhh3}, \ref{5susyqm2}
and \ref{1susyqm3})
\begin{equation}
\frac{L(L+\rm 1)}{\rm 2 \mit r^{2}} - \frac{\gamma}{r} + \frac{\rm 1}{\rm 2}
\left( \frac{\gamma}{L+\rm 1}  \right)^2 =
\frac{1}{2} \left[  W_{L}^{2} - \frac{dW_{L}}{dr}\right]
\end{equation}
whose solution is
\begin{equation}
W_{L} = -\frac{L+1}{r} + \frac{\gamma}{L+1}\;.
\label{6susyhh3}
\end{equation}
Equation \ref{6susyhh3} and \ref{6susyqm2} yield
\begin{equation}
A^{\pm}_{L} = \frac{1}{\sqrt{2}} \left( \mp \frac{d}{dr} -\frac{L+1}{r} +
\frac{\gamma}{L+1} \right).
\label{1susyhh4}
\end{equation}

{}From equation \ref{1susyhh4} and \ref{2susyqm3}, we construct the
SUSY-partner hamiltonian of $\cal H_{\mit L}$ in equation \ref{4susyhh3},
\begin{equation}
\cal H_{\mit L + \rm 1} = \mit A^{-}_{L} A^{+}_{L} =
\mit -\frac{\rm 1}{\rm 2} \frac{d^2}{dr^2} + \frac{(L + \rm 1)(\mit L+\rm
2)}{\rm 2 \mit r^{2}} - \frac{\gamma}{r} + \frac{\rm 1}{\rm 2} \left(
\frac{\gamma}{L+\rm 1}  \right)^2\; .
\label{2susyhh4}
\end{equation}

Comparing equations \ref{2susyhh3} and \ref{4susyhh3} and with equation
\ref{2susyhh4}, we realize that
\begin{equation}
\cal H_{\mit L} = \mit H_{L} + \frac{\rm 1}{\rm 2} \left( \frac{\gamma}{L+\rm
1}  \right)^2
\label{3susyhh4}
\end{equation}
\begin{equation}
\cal H_{\mit L + \rm 1} = \mit H_{L + \rm 1} + \frac{\rm 1}{\rm 2} \left(
\frac{\gamma}{L+\rm 1}  \right)^2\; .
\label{4susyhh4}
\end{equation}

\begin{figure}[htbp]
\begin{center}
\begin{picture}(400,200)
                            \put(37.5,173){\vdots}
        \put(167.5,173){\vdots}
\put(297.5,173){\vdots}
                            \put(37.5,160){\vdots}
        \put(167.5,160){\vdots}
\put(297.5,160){\vdots}
                            \put(0,130){\line(75,0){75}}
        \put(130,130){\line(75,0){75}}
\put(260,130){\line(75,0){75}}
                            \put(0,75){\line(75,0){75}}
        \put (130,75){\line(75,0){75}}
                           \put(0,20){\line(75,0){75}}
                            \put(30,0){$\cal H_{\mit \left| M \right|}$}
                        \put(155,0){$\cal H_{\mit \left| M \right| + \rm 1}$}
                              \put(280,0){$\cal H_{\mit \left| M \right|  + \rm
2}$}           \put(410,0){\ldots \ldots}
\end{picture}
\end{center}
\caption{The energy states of the hierarchy of SUSY-partner hamiltonians from
the Hartmann potential}
\label{fig2susyhh4}
\end{figure}
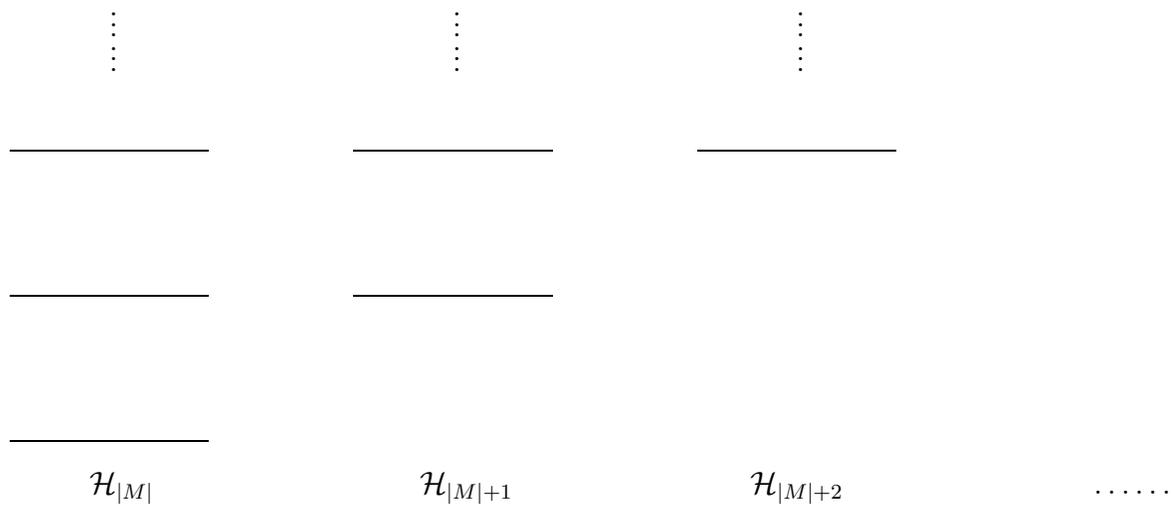

Let us now start to build up the radial eigenfunctions and in the process get
the eigenvalues. Given an $\left| M \right|$ value, the lowest $L$ value
is $L = \left| M \right|$, as can be seen in equation \ref{6susyhh2}. It is
apparent from equations \ref{3susyhh4} and \ref{4susyhh4} that we can build the
states of the hierarchy of hamiltonians as in figure \ref{fig1susyqm5}. This is
illustrated in figure \ref{fig2susyhh4}.

Since $\cal H_{\mit \left| M \right|}$ ($\cal H_{\mit \left| M \right| + \rm
1}$) and $H_{L}$ ($H_{L+1}$) differ only by a constant, (see equations
\ref{3susyhh4} and \ref{4susyhh4}) every eigenfunction of $\cal H_{\mit \left|
M \right|}$ ($\cal H_{\mit \left| M \right| + \rm 1}$) will be an eigenfunction
of $H_{L}$ ($H_{L+1}$). Hence, all we have to do is to solve for the
eigenfunctions of $\cal H_{\mit L}$. The actual energy for $H_{L}$ can be found
by letting
$H_{L}$ act on the eigenfunctions of $\cal H_{\mit L}$.

For an arbitrary $L$, equations \ref{5susyqm4} and \ref{1susyhh4} give, for the
ground states of $\cal H_{\mit L}$, $\psi^{0}_{(L)}$, the first order
differential
equation
\begin{equation}
\frac{1}{\sqrt{2}} \left(  \frac{d}{dr} -\frac{L+1}{r} + \frac{\gamma}{L+1}
\right) \psi^{0}_{(L)} = 0
\label{45p}
\end{equation}
which can easily be solved as
\begin{equation}
\psi^{0}_{(L)} = \cal N_{\mit L} \; \mit r^{L+ \rm 1} exp(-\kappa_{L} r)
\label{1susyhh5}
\end{equation}
where
\begin{equation}
\kappa_{L} \equiv \frac{\gamma}{L+1}\;.
\label{2susyhh5}
\end{equation}

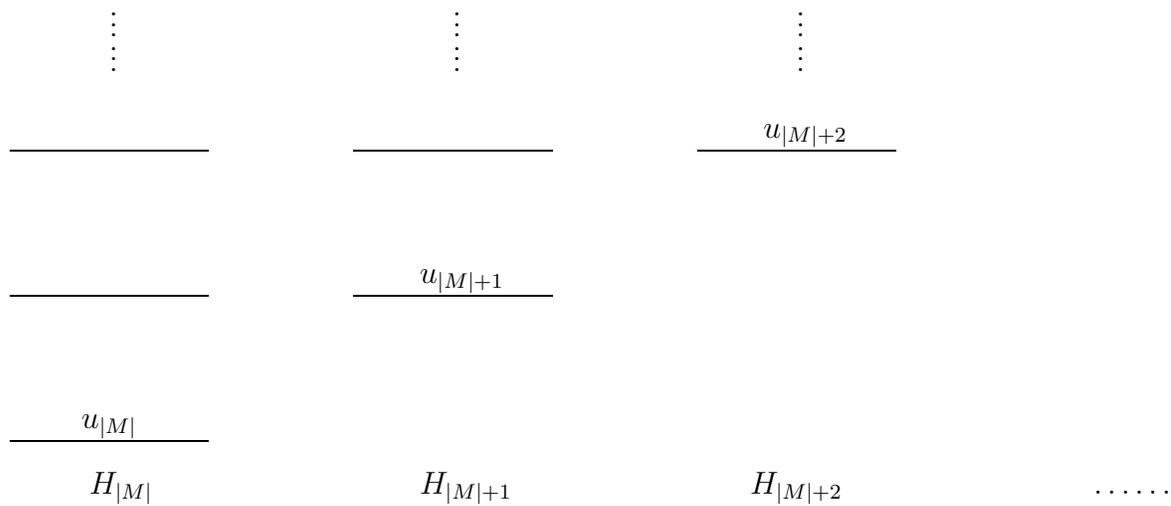
\begin{figure}[htbp]
\begin{center}
\begin{picture}(400,200)
                            \put(37.5,173){\vdots}
        \put(167.5,173){\vdots}
\put(297.5,173){\vdots}
                            \put(37.5,160){\vdots}
        \put(167.5,160){\vdots}
\put(297.5,160){\vdots}

\put(285,135){$u_{\left| M \right| + 2}$}
                            \put(0,130){\line(75,0){75}}
        \put(130,130){\line(75,0){75}}
\put(260,130){\line(75,0){75}}

        \put(155,80){$u_{\left| M \right| + 1}$}
                            \put(0,75){\line(75,0){75}}
        \put (130,75){\line(75,0){75}}
                            \put(27,25){$u_{\left| M \right|}$}
                            \put(0,20){\line(75,0){75}}
                            \put(30,0){$H_{\left| M \right|}$}
              \put(155,0){$H_{\left| M\right| + 1}$}
     \put(280,0){$ H_{\left| M \right| +  2 }$}           \put(410,0){\ldots
\ldots}
\end{picture}
\end{center}
\caption{The hierarchy of hamiltonians of the Hartmann potential and their
ground states. The $H_{L}$ here are the actual radial hamiltonian for a
particular $L$ value.}
\label{fig3susyhh6}
\end{figure}

Since $L$ is arbitrary here, we realize that equation \ref{1susyhh5} is the
expression for the eigenfunction for the lowest rung
(i.e. the ground state) of the tower of states for each of the
hamiltonians in figure \ref{fig2susyhh4}. Since they are also eigenfunctions of
$H_{L}$, we can write
\begin{equation}
u_{L} = \cal N_{\mit L} \;  \mit r^{L+ \rm 1} exp(-\kappa_{L} r)
\label{3susyhh5}
\end{equation}
where we illustrate them in figure \ref{fig3susyhh6}.

\begin{figure}[htbp]
\begin{center}
\begin{picture}(400,200)
                            \put(37.5,173){\vdots}
        \put(167.5,173){\vdots}
\put(297.5,173){\vdots}
                            \put(37.5,160){\vdots}
        \put(167.5,160){\vdots}
\put(297.5,160){\vdots}

\put(285,135){$u_{\left| M \right| + 2}$}
\put(-40,130){$\frac{-\Lambda}{\left( \left| M \right| +3 \right)^2}$}
                  \put(0,130){\line(75,0){75}}
\put(130,130){\line(75,0){75}}
\put(260,130){\line(75,0){75}}

        \put(155,80){$u_{\left| M \right| + 1}$}
\put(-40,75){$\frac{-\Lambda}{\left( \left| M \right| +2 \right)^2}$}
                 \put(0,75){\line(75,0){75}}
\put (130,75){\line(75,0){75}}
                            \put(27,25){$u_{\left| M \right|}$}
\put(-40,20){$\frac{-\Lambda}{\left( \left| M \right| +1 \right)^2}$}
                 \put(0,20){\line(75,0){75}}
                            \put(30,0){$H_{\left| M \right|}$}
              \put(155,0){$H_{\left| M\right| + 1}$}
     \put(280,0){$ H_{\left| M \right| +  2}$}           \put(410,0){\ldots
\ldots}
\end{picture}
\end{center}
\caption{Figure 3 with the energy levels labelled.}
\label{fig4susyhh6}
\end{figure}
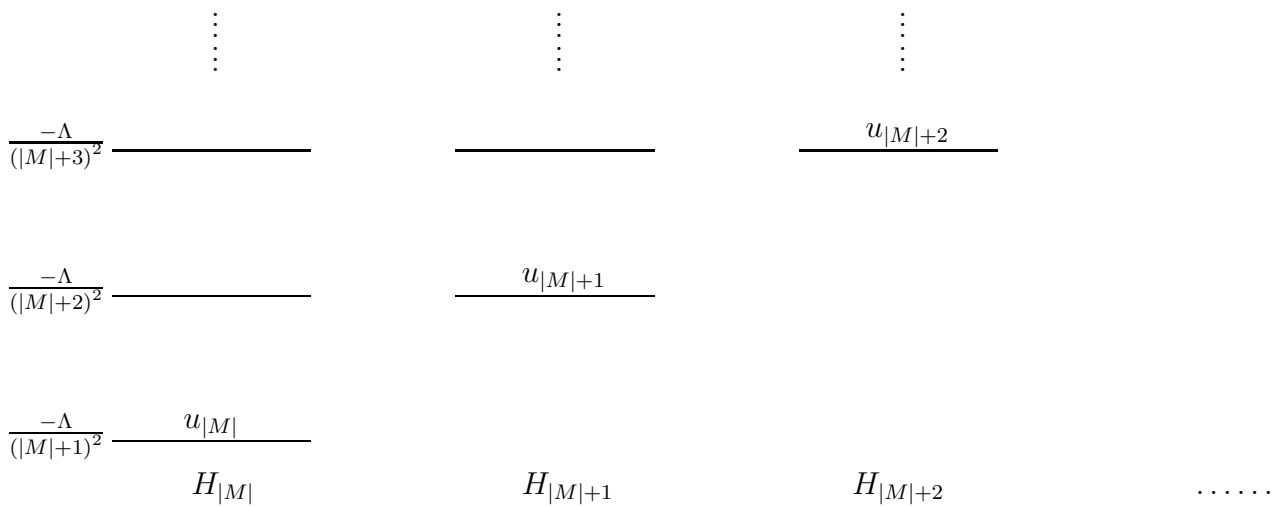

To get the actual energy, we let $H_{L}$ of equation \ref{2susyhh3} act on
equation \ref{3susyhh5}.
\begin{equation}
H_{L} u_{L} = \left[ -\frac{1}{2} \frac{d^2}{dr^2} + \left( \frac{L(L+1)}{2
r^2} - \frac{\gamma}{r} \right) \right] u_{L} = \frac{\mu E_{L}}{\hbar^2}
u_{L}\;.
\end{equation}
After some simplification, this yields
\begin{equation}
E_{L} = -\frac{\Lambda}{\left( L+1 \right)^2}\;\;\;\;\; \Lambda = \eta^2
\sigma^4 \left| \epsilon_{0} \right|.
\label{2susyhh6}
\end{equation}
We characterize the energy by the $L$ quantum number for the moment. From
equation \ref{2susyhh6}, we can label the energy levels of figure
\ref{fig3susyhh6}
as in figure \ref{fig4susyhh6}.

It is apparent from figure \ref{fig4susyhh6} that we have to label the
solutions $u$ as $u_{\left| M \right| + 1,\, \left| M \right|}$; $u_{\left| M
\right| + 2,\, \left| M \right| + 1}$;
$u_{\left| M \right| + 3,\, \left| M \right| + 2}$; \ldots due to the energy of
the states. Knowing the eigenstates at the lowest rung of the hierarchy of
hamiltonians, $u_{\left| M \right| + 1,\, \left| M \right|}$; $u_{\left| M
\right| + 2,\, \left| M \right| + 1}$;
$u_{\left| M \right| + 3,\, \left| M \right| + 2}$; etc., we can determine the
other states by the action of $A^{+}_{L}$ on these eigenstates
as in figure \ref{fig1susyqm5}. This is illustrated in figure
\ref{fig5susyhh7}.

Note that, for instance,  $u_{\left| M \right| + 3, \left| M \right|}$;
$u_{\left| M \right| + 3, \left| M \right| + 1}$; $u_{\left| M \right| + 3,
\left| M \right| + 2}$; \ldots
have the same energy $-\frac{\Lambda}{\left( \left| M  \right| + 3 \right)^2}$,
and similarly for other states at the same energy level. It is then evident
that given
$N \geq \left| M \right| + 1$, $u_{N, N-1}$; $u_{N, N-2}$; $u_{N, N-3}$;
\ldots; $u_{N, \left| M \right|}$ will all have the same energy
$-\frac{\Lambda}{N^2}$.
Hence, we can say that
\begin{equation}
E_{N} = -\frac{\Lambda}{N^2},\;\;\;\;\; \Lambda = \eta^2 \sigma^4 \left|
\epsilon_{0} \right|,\;\;\;\;\; N = L + 1 + n^{\prime},\;\;\;\;\; n^{\prime} =
0, 1, 2, \ldots
\label{1susyhh7}
\end{equation}
which means that the energy is actually labelled by $N$ and not $L$. Equation
\ref{1susyhh7} agrees with reference \cite{hart}.

{}From figure \ref{fig5susyhh7}, equations \ref{2susyhh5}, \ref{3susyhh5},
\ref{1susyhh3} and \ref{1susyhh4}, it can be shown that
\begin{equation}
R_{\left| M \right| + 1, \left| M \right|}(r) = \left[ \frac{2 \gamma}{\left| M
\right| + 1} \right]^{\left| M \right| + 3/2}
\left[ \frac{1}{\Gamma \left( 2 \left| M \right| + 3 \right)}  \right]^{1/2}
r^{\left| M \right|} e^{-\gamma r/\left[ \left| M \right| + 1  \right]}
\label{1susyhh8}
\end{equation}
\begin{equation}
R_{\left| M \right| + 2, \left| M \right| + 1}(r) = \left[ \frac{2
\gamma}{\left| M \right| + 2} \right]^{\left| M \right| + 5/2}
\left[ \frac{1}{\Gamma \left( 2 \left| M \right| + 5 \right)}  \right]^{1/2}
r^{\left| M \right| + 1} e^{-\gamma r/\left[ \left| M \right| + 2  \right]}
\label{2susyhh8}
\end{equation}
\begin{equation}
\begin{array}{lll}
R_{\left| M \right| + 2, \left| M \right|}(r) & = & -\left[ \frac{2
\gamma}{\left| M \right| + 2} \right]^{\left| M \right| + 3/2}
\left[ \frac{1}{2 \left( \left| M \right| + 2 \right) \Gamma \left( 2 \left| M
\right| + 3 \right)}  \right]^{1/2}
r^{\left| M \right|} e^{-\gamma r/\left[ \left| M \right| + 2  \right]} \\
&  & \times \left[ 2 \left| M \right| + 2 - \frac{2 \gamma r}{ \left| M \right|
+ 2  } \right]
\end{array}
\label{3susyhh8}
\end{equation}
etc. where $R_{NL}$ are normalized by
\begin{equation}
1 = \int_{0}^{\infty} \left| R_{NL} \right|^2 r^2 dr
\label{4susyhh8}
\end{equation}
and the gamma function properties used are \cite{arf}
\begin{equation}
\Gamma (z) = \int_{0}^{\infty} e^{-t} t^{z-1} dt; \;\;\;\;\; \Gamma(z+1) = z
\Gamma(z)\;.
\label{5susyhh8}
\end{equation}
Equations \ref{1susyhh8}, \ref{2susyhh8} and \ref{3susyhh8} agree with the
normalized $R_{NL}$ of reference \cite{schuch}. It is obvious that we can
eventually get all the expressions of
$R_{NL}$.

To illustrate the procedure more concretely, let us show how we can get
$R_{\left| M \right| + 2, \left| M \right|}(r)$. From figure \ref{fig5susyhh7},
\begin{equation}
u_{\left| M \right| + 2, \left| M \right|} \sim A^{+}_{\left| M \right|}\;
u_{\left| M \right| + 2, \left| M \right| + 1}.
\footnote{ Note that in equation \ref{1susyhh9} and in some of the following
equations, we use $\sim$ instead of $=$ since the procedure outlined here
cannot automatically normalize
the eigenfunctions.  At the end, we normalize the eigenfunctions using equation
\ref{4susyhh8}. }
\label{1susyhh9}
\end{equation}
Since $u_{\left| M \right| + 2, \left| M \right| + 1}$ is at the lowest rung of
$H_{\left| M \right| + 1}$, we can use equations \ref{2susyhh5} and
\ref{3susyhh5} which give
\begin{equation}
u_{\left| M \right| + 2, \left| M \right| + 1} \sim r^{\left| M \right| + 2}
e^{-\gamma r/\left[ \left| M \right| + 2  \right]}\;.
\label{2susyhh9}
\end{equation}
{}From equations \ref{1susyhh4}, \ref{1susyhh9} and \ref{2susyhh9}, we get
\begin{equation}
u_{\left| M \right| + 2, \left| M \right|} \sim \left( - \frac{d}{dr}
-\frac{\left| M \right|+1}{r} + \frac{\gamma}{\left| M \right|+1} \right)
r^{\left| M \right| + 2} e^{-\gamma r/\left[ \left| M \right| + 2  \right]}
\end{equation}
which leads to
\begin{equation}
u_{\left| M  \right| +2, \left| M  \right|} \sim r e^{-\gamma r/ \left[ \left|
M  \right| + 2 \right]}
\left( -r^{\left| M  \right|} + \frac{\gamma r^{\left| M  \right|
+1}}{\left(\left| M  \right| +2 \right) \left(\left| M  \right| +1 \right)}
\right).
\label{3susyhh9}
\end{equation}
Using equations \ref{3susyhh9} and \ref{1susyhh3} and rearranging terms, we get
\begin{equation}
R_{\left| M  \right| +2, \left| M  \right|} \sim  -e^{-\gamma r/ \left[ \left|
M  \right| + 2 \right]}
\;r^{\left| M  \right|} \left( 2 + 2 \left| M  \right| - \frac{2 \gamma
r}{\left| M  \right| +2}
\right)
\end{equation}
or
\begin{equation}
R_{\left| M  \right| +2, \left| M  \right|} = - \cal N \mit  e^{-\gamma r/
\left[ \left| M  \right| + \rm 2 \right]}
\;r^{\left| M  \right|} \left( \rm 2 + 2 \mit \left| M  \right| - \frac{\rm 2
\mit \gamma r}{\left| M  \right| +\rm 2}
\right)
\label{5susyhh9}
\end{equation}
where $\cal N$ is the normalization constant. Normalizing equation
\ref{5susyhh9} using equations \ref{4susyhh8} and \ref{5susyhh8} leads to
equation \ref{3susyhh8}.
\clearpage

\section{Conclusion}
\label{concl}
\indent

The ring-shaped Hartmann potential was first introduced in 1972 \cite{hart}. It
has been used to describe ring-shaped molecules like the benzene molecule in
theoretical chemistry \cite{schuch}. In an attempt to introduce the methods of
supersymmetric quantum mechanics  in
quantum chemistry, we obtain the eigenvalues and radial eigenfunctions of the
Hartmann potential in spherical coordinates using SUSYQM
techniques.

A key result in SUSYQM is the intimate relationship of the eigenvalues and
eigenfunctions of the hierarchy of SUSY-partner hamiltonians.
This can be very useful in solving the Schr\"{o}dinger equation of a
complicated hamiltonian if its SUSY-partner hamiltonian is easily solvable.

A very useful result in the present discussion is the fact that $A^{-} \psi^{0}
= 0$. This enabled us to solve a first order differential
equation (as in equation \ref{45p}) instead of the second order Schr\"{o}dinger
differential equation to give us the eigenfunctions and eigenvalues of the
states
at the lowest rung of the tower of states of each of the hamiltonians in the
hierarchy. The rest of the eigenfunctions and eigenvalues
are then easily computed by applying the corresponding $A^{+}_{L}$ operators to
these eigenfunctions.

As already indicated in this paper, the separated equations of the Hartmann
potential and the hydrogen atom greatly resemble each other.
In fact, the technique employed here was first applied to the hydrogen atom
\cite{schwabl}. A number of studies of the SUSY features of the
Coulomb problem in hydrogenic atoms have been made over the past years
\cite{bluhm,kost,nieto,hay}. These studies may very well lead to
some further insights into the workings of SUSY in the Hartmann potential due
to the similarity of its separated equations with that of
the hydrogen atom.

The preceding paragraph actually led the author to believe that if the Hartmann
problem is treated using the SUSY formulation of Haymaker
and Rau \cite{hay}, the eigenstates of different values of $\eta \sigma^2$ can
be related. This will be developed in a future publication.

With the above considerations, the author believes that quantum chemistry is a
field where the concepts and techniques of SUSYQM can be put
to good use.
\clearpage

\section*{Figure Captions}
\begin{enumerate}
   \item The hierarchy of hamiltonians and the action of the operators
$A^{\pm}_{n}$ on the degenerate eigenstates
   \item The energy states of the hierarchy of SUSY-partner hamiltonians from
the Hartmann potential
   \item The hierarchy of hamiltonians of the Hartmann potential and their
ground states. The $H_{L}$ here are the actual radial hamiltonian for a
particular $L$ value.
   \item Figure 3 with the energy levels labelled.
   \item The energy eigenstates of the Hartmann potential. The action of the
$A^{+}_{L}$ operators are explicitly shown to indicate how the other states are
obtained from the states at the lowest rung of the hierarchy of hamiltonians.
\end{enumerate}
\clearpage

\clearpage
\clearpage
\clearpage
\clearpage
\begin{figure}[htbp]
\begin{center}
\begin{picture}(400,200)

\put(37.5,173){\vdots}
                  \put(167.5,173){\vdots}
                                               \put(297.5,173){\vdots}

\put(37.5,160){\vdots}
                  \put(167.5,160){\vdots}
                                               \put(297.5,160){\vdots}

                                                 \put(95,145){$A^{+}_{\left| M
\right|}$}
\put(215,145){$A^{+}_{\left| M \right| + 1}$}

\put(130,135){\vector(-1,0){55}}
                        \put(260,135){\vector(-1,0){55}}

\put(15,135){$u_{\left| M \right| + 3, \left| M \right|}$}
                  \put(140,135){$u_{\left| M \right| + 3, \left| M \right| +
1}$}                                              \put(275,135){$u_{\left| M
\right| + 3, \left| M \right| + 2}$}
\put(-40,130){$\frac{-\Lambda}{\left( \left| M \right| +3 \right)^2}$}
\put(0,130){\line(75,0){75}}
                  \put(130,130){\line(75,0){75}}
                                               \put(260,130){\line(75,0){75}}

                                                 \put(95,90){$A^{+}_{\left| M
\right|}$}

\put(130,80){\vector(-1,0){55}}

\put(10,80){$u_{\left| M \right| + 2, \left| M \right|}$}
                  \put(140,80){$u_{\left| M \right| + 2, \left| M \right| +
1}$}
\put(-40,75){$\frac{-\Lambda}{\left( \left| M \right| +2 \right)^2}$}
\put(0,75){\line(75,0){75}}
                  \put (130,75){\line(75,0){75}}

\put(10,25){$u_{\left| M \right| + 1, \left| M \right|}$}
\put(-40,20){$\frac{-\Lambda}{\left( \left| M \right| +1 \right)^2}$}
\put(0,20){\line(75,0){75}}

\put(30,0){$H_{\left| M \right|}$}
                  \put(155,0){$H_{\left| M\right| + 1}$}
                                              \put(280,0){$ H_{\left| M \right|
+  2}$}  \put(410,0){\ldots \ldots}
\end{picture}
\end{center}
\caption{The energy eigenstates of the Hartmann potential. The action of the
$A^{+}_{L}$ operators are explicitly shown to indicate how the other states are
obtained from the states at the lowest rung of the hierarchy of hamiltonians.}
\label{fig5susyhh7}
\end{figure}
\clearpage

\label{ref}

\end{document}